\documentclass[fleqn,usenatbib]{mnras}

\usepackage{amssymb}
\usepackage{amsmath}
\usepackage{graphicx}
\usepackage{float}
\usepackage{nccmath}
\usepackage{dblfloatfix} 
\pdfoutput=1

    \title{Can the orbital distribution of Neptune's 3:2 mean motion resonance result from stability sculpting?}
    
    \author[Balaji et al.]{S. Balaji$^{1}$,
            N. Zaveri$^{1}$,
            N. Hayashi$^{1}$,
            A. Hermosillo Ruiz$^{1}$,
            J. Barnes$^{1,3}$,
            \newauthor
            R. Murray-Clay$^{1}$,
            K. Volk$^{2}$,
            J. Gerhardt$^{1}$,
            and Z. Syed$^{1}$
          \\
   $^{1}$University of California Santa Cruz (UCSC), Department of Astronomy and Astrophysics, 1123 High St, Santa Cruz, CA 95060 \\
    $^{2}$Lunar and Planetary Laboratory, The University of Arizona, 1629 E University Blvd, Tucson, AZ 85721 \\
    $^{3}$Northwestern University, Department of Physics and Astronomy, Evanston, IL 60208
    }

    \pubyear{2020}
\begin{document}
\raggedbottom
\label{firstpage}
\pagerange{\pageref{firstpage}--\pageref{lastpage}}
\maketitle

  \begin{abstract}
        We explore a simplified model of the outcome of an early outer Solar System gravitational upheaval during which objects were captured into Neptune's 3:2 mean motion resonance via scattering rather than smooth planetary migration. 
        We use N-body simulations containing the Sun, the four giant planets, and test particles in the 3:2 resonance to determine whether long-term stability sculpting over 4.5 Gyr can reproduce the observed 3:2 redresonant population from an initially randomly scattered 3:2 population.
        After passing our simulated 3:2 resonant objects through a survey simulator, we find that the semimajor axis ($a$) and eccentricity ($e$) distributions are consistent with the observational data (assuming an absolute magnitude distribution constrained by prior studies), suggesting that these could be a result of stability sculpting. However, the inclination ($i$) distribution cannot be produced by stability sculpting and thus must result from a distinct process that excited the inclinations. Our simulations modestly under-predict the number of objects with high libration amplitudes ($A_\phi$), possibly because we do not model transient sticking. Finally, our model under-populates the Kozai subresonance compared to both observations and to smooth migration models.  Future work is needed to determine whether smooth migration occurring as Neptune's eccentricity damped to its current value can resolve this discrepancy.  
        
  \end{abstract}

  \maketitle

%

\section{Introduction}\label{intro}

    The dynamical structure of small bodies in the Solar System's trans-Neptunian region indicates that the system's ice giants formed closer to the Sun than they orbit today. In particular, the large population of trans-Neptunian objects (TNOs) detected in mean motion resonances with Neptune suggests that early in its lifetime, Neptune either migrated outward from a closer-in orbit due to angular momentum transfer with nearby planetesimal debris or was dynamically scattered due to interactions with the other giant planets (or both; for reviews see, e.g., \citealt{Morbidelli:2008,Nesvorny:2018, morbidelli2020, Gomes:2018}). 
    Recent results from well-characterized surveys of the trans-Neptunian region have enabled direct comparisons between these models and the distribution of observed resonant orbits.   
    In this paper, we investigate whether the observed orbital distribution of TNOs in the 3:2 mean motion resonance (MMR) with Neptune is consistent with the class of models in which Neptune is dynamically scattered.  To do so, we test whether this population can be produced by an initially scattered population of TNOs for which no preferential resonance capture has occurred, which is then sculpted over the age of the Solar System as unstable objects are lost. We refer to this process as ``stability sculpting."

    The nature of the Solar System's early dynamical evolution is still uncertain, but two end-member models are often discussed: gravitational upheaval and smooth migration. 
    Both have a similar pre-evolution state, with all of the giant planets on nearly-circular, co-planar orbits
    with semi-major axes interior to Neptune's current orbit  and an initial massive planetesimal disk extending from the giant planet region to roughly 34~au
    (see, e.g. \citealt{Levison2008}; though at least some low-mass portion of the disk also extended out to include the current cold classical population at $\sim$45~au as discussed in, e.g., \citealt{Mckinnon:2020,Gladman:2021}). 
    The two models differ in their implications for how Neptune's exterior mean motion resonances are filled. 
    In the most violent upheaval models, the giant planets have direct gravitational interactions that scatter Neptune nearly directly to its current location (see, e.g., \citealt{Tsiganis:2005, Gomes2005,deSousa:2020}; see also reviews by \citealt{Morbidelli:2008,Nesvorny:2018,morbidelli2020}).
    In this type of scenario, most of the planetesimals are strongly scattered with some landing at random in the final locations of Neptune's mean motion resonances \citep[e.g.,][]{Levison2008,pike2017}.
    Smooth migration models are characterized by a slower, gradual outward migration of the planets, during which planetesimals are captured into resonant orbits as the locations of the resonances sweep past them \citep[e.g.,][]{malhotra1993,malhotra1995,Hahn:2005}.

    \begin{figure*}
        \centering
        \includegraphics[scale=0.46]{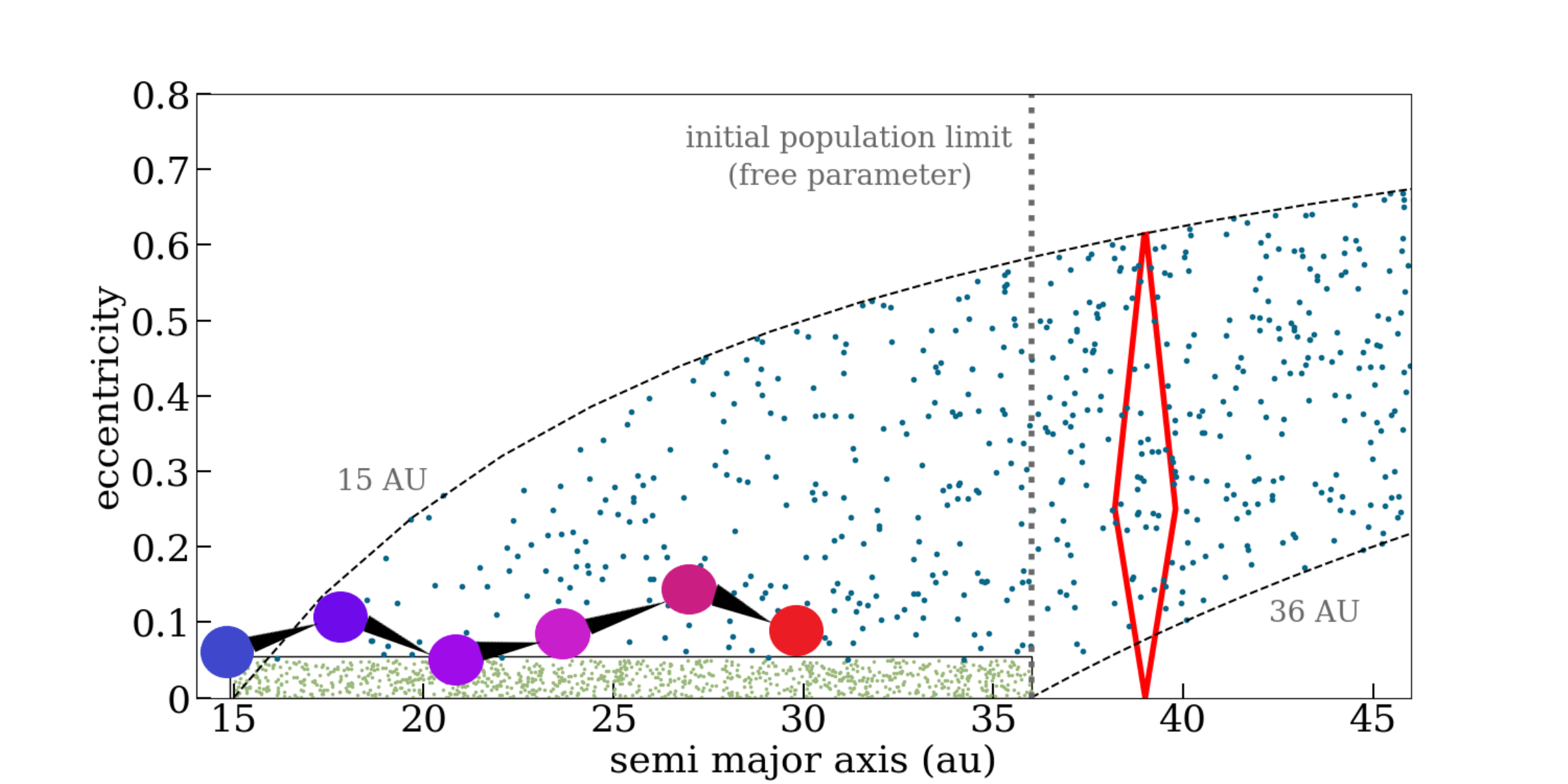}
        \caption{A schematic representation of a scattering origin for the 3:2 resonant population, providing the motivation for our simulation initial conditions. The giant planets and the objects that make up today's 3:2 resonant population were initially closer to the Sun before the giant planets migrated to their current orbits. The initial disk of TNOs was dynamically cold (dense green dots) but  scattered stochastically onto dynamically excited orbits (blue dots) due to a gravitational upheaval amongst the giant planets. Some of these TNOs randomly land in the area of phase space (red diamond) where the 3:2 resonance is currently located. The radial extent of the initial cold disk of TNOs that experiences scattering is a free parameter in our model, manifesting as a maximum initial perihelion distance of 36 au (discussed in Section \ref{discussion} As the TNOs scatter, Neptune does a random walk to get to its current position, shown by the large circles from 15 au through 30 au. We assume Neptune's displacement happens fast enough such that it 'appears' at its current position, thus allowing for the particles to be at any libration amplitude at the start of the N-body simulation.}
        
         \label{fig:scattering}
     \end{figure*}

    In gravitational upheaval models, the ice giants exhibit chaotic orbital evolution, meaning that their final orbits are not easily controlled in N-body simulations. It is thus computationally challenging to perform pure upheaval simulations suitable for high fidelity comparisons with observations of resonant TNOs.  Our aim in this paper is to sidestep this challenge by testing a generalized model of the outcome of a gravitational upheaval scenario, including long-term sculpting by dynamical instabilities.
    We assume a simplified scenario where gravitational perturbations in the early Solar System scattered or ``kicked" trans-Neptunian planetesimals onto various orbits beyond Neptune's current semi-major axis.
    The giant planets simultaneously undergo strong mutual perturbations, including scattering events, that cause them to spread out.
    Once the giant planets arrive at and settle into their current, stable orbits, some of those scattered planetesimals will remain in stable/meta-stable orbits. 
    These remaining TNOs are categorized into different dynamical sub-populations (see, e.g., \citealt{gladman2008}).

    To test a simplified model of a giant planet dynamical upheaval, here we focus on the dynamical evolution of the 3:2 MMR population, which is located at a semimajor axis $a=39.4$~au. Our reason for focusing on this population stems from two key points: 
    \begin{enumerate}[I.]
    
        \item There is a significant characterized observational sample of the 3:2 MMR population from multiple well-characterized surveys \citep{petit2011, alexandersen2016, bannister2016, bannister2018}. 
        The Outer Solar System Origins Survey ensemble (OSSOS+) is a compilation of these surveys that contains field pointings, field depths, and tracking fractions at different magnitudes and on-sky rates of motion that can be combined with the OSSOS survey simulator to provide robust comparisons between models and observations (see, e.g., \citealt{Lawler:2018ss}).

        \vspace{\baselineskip}
        
        \item The 3:2 MMR population is also an ideal population to study long-term stability due to the fact that it is a strong first-order resonance. 
        The resonance hosts enough stable phase space that different emplacement mechanisms may have populated the resonance in observationally distinguishable ways.  
    
     \end{enumerate}
    

    Our work uses a simplified model of the outcome of a planetary upheaval scenario rather than direct simulations of the giant planets' early evolution to avoid the numerical complications presented by including the strong planet-planet interactions that occur during the actual epoch of planetary migration/upheaval. 
    \citealt{volk2019} highlights the difficulty in producing reasonable statistics for the final distributions of outwardly scattered planetesimals in smooth migration simulations. 
    Even without planet-planet close encounters, the interactions between planets during migration introduce significant randomness to the planet outcomes; coupled with the very low efficiency at which test particles land on even meta-stable orbits in regions of interest such as the present-day 3:2 resonance, it becomes computationally challenging to produce statistically meaningful resonant populations.
    When even stronger planet-planet interactions are introduced, the numerical challenges in finding simulation initial conditions that result in well-behaved final giant planet orbits \textit{and} then integrating them with enough test particles to result in a sufficiently large final 3:2 population are dramatically magnified. We discuss this further in Section \ref{validation}.

    No two simulations of giant planet instabilities are exactly alike, and the precise distribution of scattered planetesimals that remain at the end of the scattering epoch may be affected by mean motion and secular resonances.  
    However, scattered planetesimals are typically roughly evenly distributed along trajectories with pericenters in the scattering region.  We therefore consider a population of objects that ``fills phase space" for different ranges of perihelion distances in the 3:2 MMR with Neptune as an approximation of the outcome of an epoch of scattering (see Section~\ref{ss:init}).
    We perform N-body simulations on a 4.5 Gyr timescale to allow the resonant phase space to be sculpted by long-term stability. 
    We can then test this modeled population against the observed 3:2 resonant population by subjecting our model to the OSSOS+ ensemble biases and comparing the simulated detections to the real ones across a variety of parameters (e.g., eccentricity $e$, inclination $i$, and resonant libration parameters).

     Section~\ref{sim} presents our model and simulation setup along with the resulting distribution of resonant objects over time. 
     Section~\ref{ossos} provides a description of how the simulation is passed into the OSSOS survey simulator to produce simulated detected objects. 
     We discuss the validity and accuracy of our model in Section~\ref{discussion} and summarize in Section~\ref{summary}.

    \section{Simulations}\label{sim}
    We conduct an N-Body simulation using the Python package \textsc{rebound} \citep{Rein:2012} with the \verb|WHFast| integrator \citep{rein2015} to mimic the evolution of the 3:2 MMR population.  
    The solar system's four giant planets are initialized with their current orbital elements and the
    TNOs are treated as massless test particles.  
    We verify that TNOs that undergo close encounters with the giant planets are quickly lost from our region of interest, justifying our choice of integrator.  
    
    To generate a sample for comparison with observational data, we fill phase space in the vicinity of the 3:2 MMR with randomly-generated test particles with uniformly-drawn pericenter distances, $q$, and semi-major axes, $a$, and then integrate for 4.5 Gyr.
    The non-resonant and thus less stable particles are  ``shaved" away over time, just leaving the stable 3:2 resonant particles. This is similar to, for example, the work of \cite{tiscareno2009} who used long-term integrations to show how the 3:2 resonant population evolves over time for a different initial population. 

Scattering outcomes show that over the limited semi-major axis range we consider, particles are distributed roughly evenly in $a$ and $q$. The particles lay along lines of constant pericenter corresponding to the region in which scattering occurs (similar assumptions were made in, e.g., the \citealt{Levison2008} model for the post-instability populations), thus influencing our initial conditions. Dynamical upheaval simulations typically end with at least a brief phase of low-eccentricity, residual migration of Neptune \citep[e.g.][]{Levison2008}, which may generate additional features in the 3:2 MMR population. We comment on this possibility in Sections \ref{libamp} and \ref{sss:kz}.

\subsection{Model Overview}
\label{modeloverview}
To construct the initial state of our simulations, we assume planetesimals are scattered outward at some early epoch and then Neptune itself is scattered outward and then damped to its current orbit on a timescale fast enough such that it can be treated (from the perspective of the previously scattered planetesimals in what is now the region of the 3:2 resonance) as ``appearing" at its current orbit with a semi-major axis of $a=30.1$~au. Thus, at the end of the planetary upheaval, the 3:2 resonances is essentially laid on top of a previously scattered population of planetesimals whose perihelia are at random phases relative to Neptune; this has the effect of more or less randomly filling the libration phase space of the resonance over a range of eccentricities set by the earlier scattering processes.See Figure \ref{fig:scattering} for a schematic describing the assumed initial scattering. 

Present-day Neptune can scatter objects with perihelia $\lesssim38$ au (see, e.g., discussion in \citealt{Gladman:2021}), and non-resonant objects with q $\lesssim 33$ au are scattered on very short timescales (see, e.g., \citealt{Tiscareno2003}). During a scattering scenario, Neptune's semi-major axis and eccentricity are unknown. For example, if Neptune had a semi-major axis of 28 au and an eccentricity of 0.2 at some point in its evolution, its apocenter was at 33.6au, and it could scatter objects with pericenters a few au more distant on short timescales.  To encompass this uncertainty within our model, we consider initial populations for which particle pericenters extend to maximum values  between 33 and 38 au.  Rather than running multiple simulations, we analyse different subsets of our initial particle distribution, with each subset representing a different outcome of the epoch of planet scattering. Figure \ref{fig:scattering} illustrates this choice through a free parameter in  perihelion distance (initial population limit), which we vary until we match observations. By finding the initial perihelion distance that provides a best fit with the data, we find a potential limit to the disk region Neptune was able to scatter during any high-eccentricity phases it might have experienced. 

\subsection{Approach Validation}
\label{validation}
\begin{figure}
         \centering
         \includegraphics[scale=0.8]{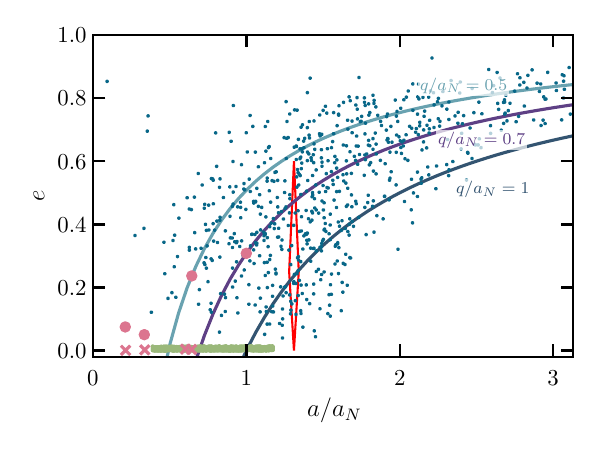}
         \caption{Ten million year snapshot of a limited-scope planetary upheaval simulation as a proof of concept for filling eccentricity and semi-major axis phase space in the 3:2 MMR. One might worry that large scattering due to the giant planets would exclude particles from the 3:2 resonance (red diamond) because resonant particles typically avoid encountering Neptune at pericenter, but this doesn't happen since Neptune jumps around substantially. The 3:2 resonant region is filled with test particles at various pericenters. 
        The initial conditions of Jupiter, Saturn, Uranus, and Neptune (pink x) were motivated by \citet{Tsiganis:2005}, where Jupiter and Saturn start near their 2:1 resonance and then undergo divergent migration. The test particles (green points) are initialized uniformly in $a$ and $e$ and $i$ from 10 to 30 au, 0 to 0.01, and 0 to 1$^\circ$ respectively. The final positions of the giant planets (pink circles) do not match today's positions, so we show the objects' semi major axis as a ratio with Neptune's final semi major position in the simulation ($a_N$ = 25.6 au). The final position of the test particles (blue points) are scattered by Neptune and Uranus, and they occupy a large amount of pericenter phase space. Three curves of constant pericenter are shown for reference, where Neptune's final apocenter is 1.3$a_N$ (33.4 au), consistent with scattering particles a few au beyond.}

         \label{fig:scatteringsim}
     \end{figure}

As a proof of concept that the simplified distribution illustrated in Figure~\ref{fig:scattering} is reasonable, we performed a very limited-scope direct simulation of a planetary upheaval scenario using the \textsc{mercurius} integrator within \textsc{REBOUND}. Similar in philosophy to the hybrid orbital integrator used by Mercury \citep{Chambers:1999}, \textsc{mercurius} combines the \textsc{whfast} and \textsc{ias15} \citep{ias15} integrators in order to follow massive bodies through mutual close-encounters. 
We used planetary initial conditions similar to those in \citet{Tsiganis:2005} and allowed the giant planets to perturb each other and a disk of massless test particles. 
    We tracked the system for 10 Myr until Neptune was scattered outward to nearly its present-day semimajor axis and the planets' orbits stabilized. 
    We then examined the distribution of outwardly scattered test particles in the vicinity of the simulated Neptune's 3:2 MMR, which is shown in Figure~\ref{fig:scatteringsim}. 
    We find that the test particles are distributed reasonably similarly to our assumed distribution described above. 
    We note that even this short, simplified simulation (we have ignored, for example, the effects of the massive planetesimal disk) required a significant amount of trial and error and hand-tuning to produce. It would require significantly more fine-tuning to produce a final Neptune orbit that acceptably matches present-day Neptune, and simulating enough test particles to fill the 3:2 resonant region is beyond our computational capabilities; this highlights why we strongly prefer our simplified approach to studying a reasonable post-upheaval distribution. 
    
\subsection{Initial conditions and resonances}\label{ss:init}
  
       \begin{figure*}
        \centering
        \includegraphics[scale=0.35]{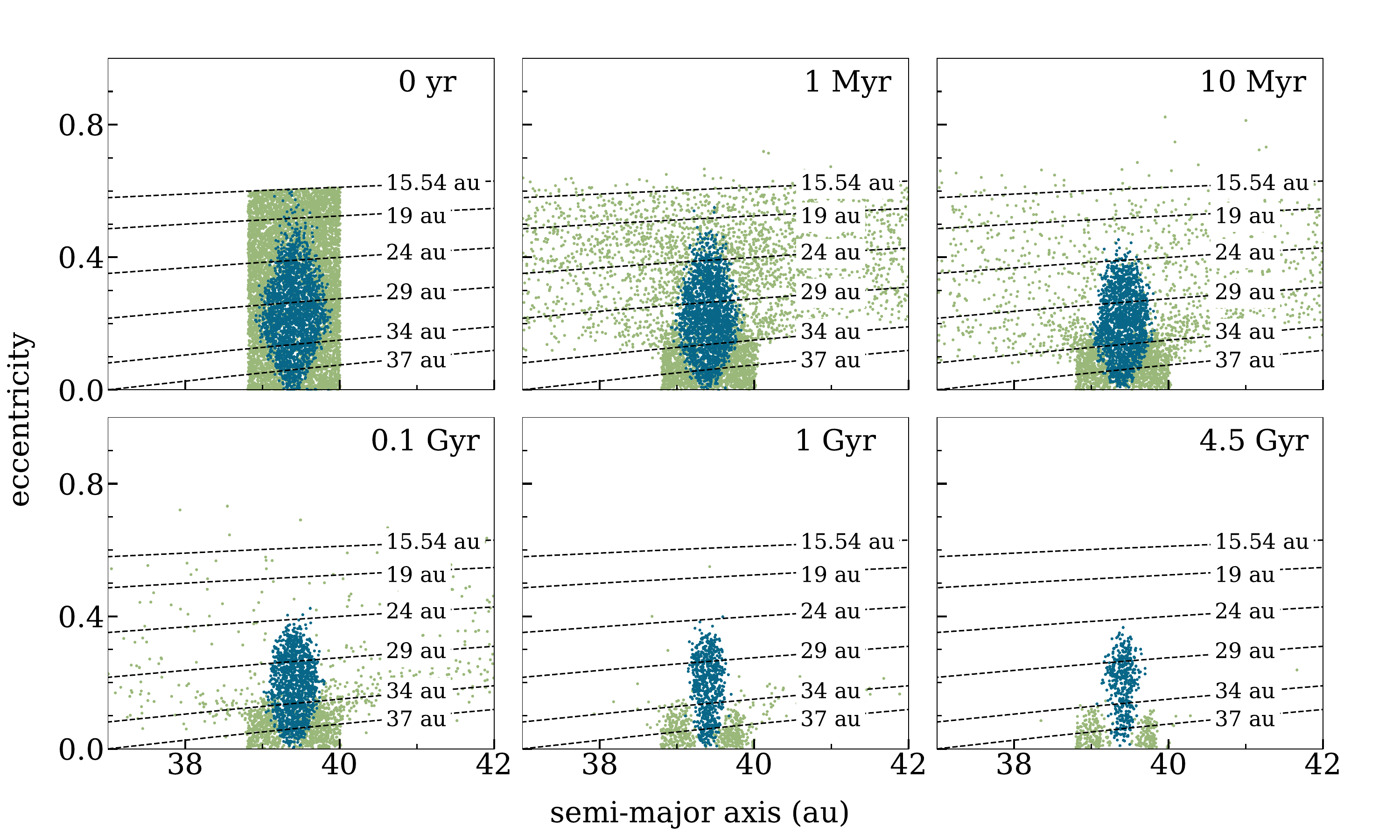}
        \caption{Over the age of the solar system, particles in Neptune's exterior 3:2 resonance (blue dots) are more dynamically stable than nearby non-resonant particles (green dots).We show eccentricity ($e$) vs semimajor axis ($a$) for these particles at six snapshots during our simulation; the particles are initialized uniformly in $a$ and pericenter distance ($q$) near the resonance (top left panel at $t=0$; see Section~\ref{ss:init}) and evolved under the influence of the Sun and four giant planets.  At each of the six displayed snapshots in time, an object is considered to be in resonance if its resonant argument ($\phi$) librates over a $10^5$ year integration started at the snapshot time. Since scattering models typically fill phase space along lines of constant pericenter, these lines (dashed) are provided for reference.  At 4.5Gyr, remaining resonant objects have $e \lesssim 0.4$ and $q \gtrsim 24$au and remaining non-resonant objects have $e \lesssim 0.15$ and $q \gtrsim 34$au.  
        }
         \label{fig:es}
     \end{figure*}
     
    Our model consists of the Sun, the four giant planets (Jupiter, Saturn, Uranus, and Neptune) and 10270  test particles that represent TNOs. 
    The giant planets are given their initial spatial parameters from NASA's JPL Horizons Ephemeris site \citep{horizons}.\footnote{Planet initial conditions were downloaded with Julian date 2458970.5 from \url{https://ssd.jpl.nasa.gov/horizons.cgi}} 
    The test particles' longitudes of ascending node ($\Omega$), arguments of pericenter ($\omega$), and mean anomalies ($M$) were randomly chosen from their full possible range, while the ranges for semi-major axis ($a$), pericenter distance ($q$), and inclination ($i$) were determined through pilot simulations (See Table \ref{initParams}). 
    
    We chose the initial range of semi-major axes to be centered around the exact resonant orbit with a wide enough range to yield a small padding of non-resonant particles on either side (see Figure \ref{fig:es}).
    In a series of pilot simulations with the initial eccentricity range set from 0-1, we found no resonant particles with eccentricity above 0.6 on a 1 Gyr timescale. 
    We therefore restrict our eccentricity range for our long simulations to $e<0.6$ for computational efficiency. 
    Upon running simulations for 1 billion years with both a uniform $e$ and uniform pericenter distance, $q = a(1-e)$, distribution, there was no notable difference between their respective time-evolved distribution in semimajor axis-eccentricity space which is most likely due to the limited $a$ range (plots not shown). Therefore, we use a uniform $q$ distribution to generate the initial eccentricity range, given our assumption that Neptune (and possibly other giant planets) kicked the planetesimals outward prior to the start of our simulations, suggesting that the objects' pericenters should be in the scattering region.

    Our pilot simulations also demonstrated that the inclination distribution of TNOs in the 3:2 MMR evolve only modestly over the lifetime of the simulation for inclinations ranging from $i=0$-90$^\circ$ (consistent with \citealt{tiscareno2009}'s finding that stability in the 3:2 resonance is not strongly affected by orbital inclination). 
    We thus assume that the emplacement mechanism, or evolution prior to emplacement, must set the current inclination distribution of the 3:2 resonance and that our initial conditions for $i$ must be similar to the current distribution (see \citealt{Li:2014_1} for an in-depth discussion).
    The initial inclination values for our test particles are randomly sampled from the differential inclination distribution modeled as $\sin i$ times a Gaussian \citep[e.g.][]{Brown:2001}. 
    When our modeled inclination distribution is compared to the observed one, the best match was a Gaussian width $\sigma_i = 14^\circ$ which is the best-fit value found for the 3:2 MMR in \cite{volk2016}.  
    
    To identify particles in the 3:2 MMR, we examine the time evolution of the particles' resonant argument, $\phi$, which is given by:
    \begin{ceqn}
    \begin{align}
        \phi = 3\lambda_{tno}-2\lambda_{N}-\varpi_{tno}, 
        \label{eq:phi}
    \end{align}
    \end{ceqn}
    where $\lambda_{tno}$ and $\lambda_N$ are the mean longitudes of the TNO and Neptune, and $\varpi_{tno}$ is the TNO's longitude of pericenter.
    The $\phi$ value of a particle in the 3:2 resonance librates around a central value of $\pi$ with a half-amplitude less than $\pi$. 
    For particles that librate within the 3:2 resonance, we also check if they are in the Kozai subresonance (sometimes also referred to as the Kozai-Lidov resonance; see, e.g., \citealt{Morbidelli:1995} for a discussion of this subresonance within the 3:2 MMR). 
    The Kozai resonance within the 3:2 resonance refers to the libration of an object's argument of pericenter, $\omega$; this corresponds physically to the location of pericenter librating around a fixed point relative to where the orbit intersects the ecliptic plane. For the 3:2 resonant particles in Kozai,  $\omega$  typically librates around a central value of either about $\frac{\pi}{2}$ or about $\frac{3\pi}{2}$.

     \begin{table}
\caption{Simulated particle initial orbital parameters. Particles are uniformly distributed in the given ranges, save for inclination which follows a modified Gaussian distribution.}
\label{initParams}

\begin{tabular}{|l|l|}
\hline
Semi-major axis, a (au) & 38.81 - 40.0 \\

Pericenter, q (au) & 15.54 - 40.0 \\

Longitude of ascending node, $\Omega$ & $0 - 2\pi$ \\
Argument of perihelion, $\omega$ & $0 - 2\pi$ \\
Mean anomaly, $M$ & $0 - 2\pi$ \\
Inclination, i (degrees) & $\frac{dN(i)}{di} \propto sin(i) exp(\frac{-i^2}{2\sigma_i^2}$) \\
Inclination width, $\sigma_i$ &  $14^\circ$ \\

\hline
\end{tabular}
\end{table}
    
    \subsection{Simulation Setup}\label{ss:sim_setup}
    
    Our integration has a total of 10270 test particles integrated for 4.5 Gyrs along with the four giant planets. 
    In an effort to be more time-efficient, we ran 158 separate simulations, each with the sun, the giant planets, and 65 test particles.  
    We confirmed that the giant planets evolved identically in each simulation. Resonance libration in the 3:2 MMR occurs on $10^4$--$10^5$-year timescales, and Kozai libration occurs on $10^6$--$10^7$-year timescales.  
    Running a 4.5 Gyr integration with thousands of test particles with frequent enough outputs to identify resonance libration generates too much data to be feasible.

    To make our simulations as time and resource efficient as possible, we split the integration into 3 parts: first is a 4.5 Gyr integration that saves snapshots at times of interest, second, a $10^5$ years integration used for determining which particles are in the 3:2 MMR at each snapshot in time, and third, a $50$ Myr integration used for determining membership in the Kozai subresonance. We set \textsc{rebound}'s internal timestep to 0.2 years, which is small enough to ensure accuracy for our simulation.
    We use the symplectic integrator \verb|whfast|, which provides a necessary increase in accuracy by averaging the total energy error at the end of the simulation and minimizes the propagation of error \citep{rein2015}.

The first integration runs for 4.5 Gyr and takes ``snapshots" of the state of the simulation at 0 years, 1 Myr, 10 Myr, 0.1 Gyr, 1 Gyr, and 4.5 Gyr. Starting from each snapshot, we use a second high-resolution $10^5$ year integration to identify resonant particles as those whose resonant argument, $\phi$, is confined to remain within the range $\phi=$5-355$^\circ$ over the typical resonant timescale. 
We can also measure the``tightness" of the resonance by finding the object's libration amplitude ($A_{\phi}$) which is defined as the half-width of the range of $\phi$. 
Operationally, $A_{\phi}$ is found by taking the difference between the maximum and minimum values of $\phi$ over $10^5$ years and dividing by 2. 
Since the libration timescale for the Kozai subresonance is significantly longer, we run a third set of integrations starting from the 0.1 Gyr, 1 Gyr, and 4.5 Gyr snapshots that run for 50 Myr and output at sufficient resolution to check for Kozai resonance. 
We consider a 3:2 resonant particle to also be in the Kozai resonance if the object's $\omega$ librates within either $\omega=$5-175$^\circ$ or $\omega=$185-355$^\circ$. 
Kozai objects can librate outside of these ranges but the above cut provide a simplified, uniform check that identifies most of the Kozai particles (see section \ref{sss:kz} for more details).

\subsection{Simulation Results}\label{ss:sim_results}

The simulation effectively ``sculpts" the 3:2 resonant population over a 4.5 Gyr period. 
Figure~\ref{fig:es} shows the eccentricity vs. semimajor axis evolution of our simulated particles; the less stable particles scatter away over time, while the most stable favor lower eccentricities and are tightly packed at the center of the resonance. 
Most of the non-resonant particles are lost on relatively short timescales, and on longer timescales resonant particles with perihelia near Uranus, ($q\approx$19~au) are lost as well because they are not phase protected from that planet. 
At 4.5 Gyr, a small, non-resonant classical population remains on either side of the 3:2 MMR; this population is further discussed in Section~\ref{sss:cls}.
    
\begin{figure}
    \centering
    \includegraphics[scale=0.16]{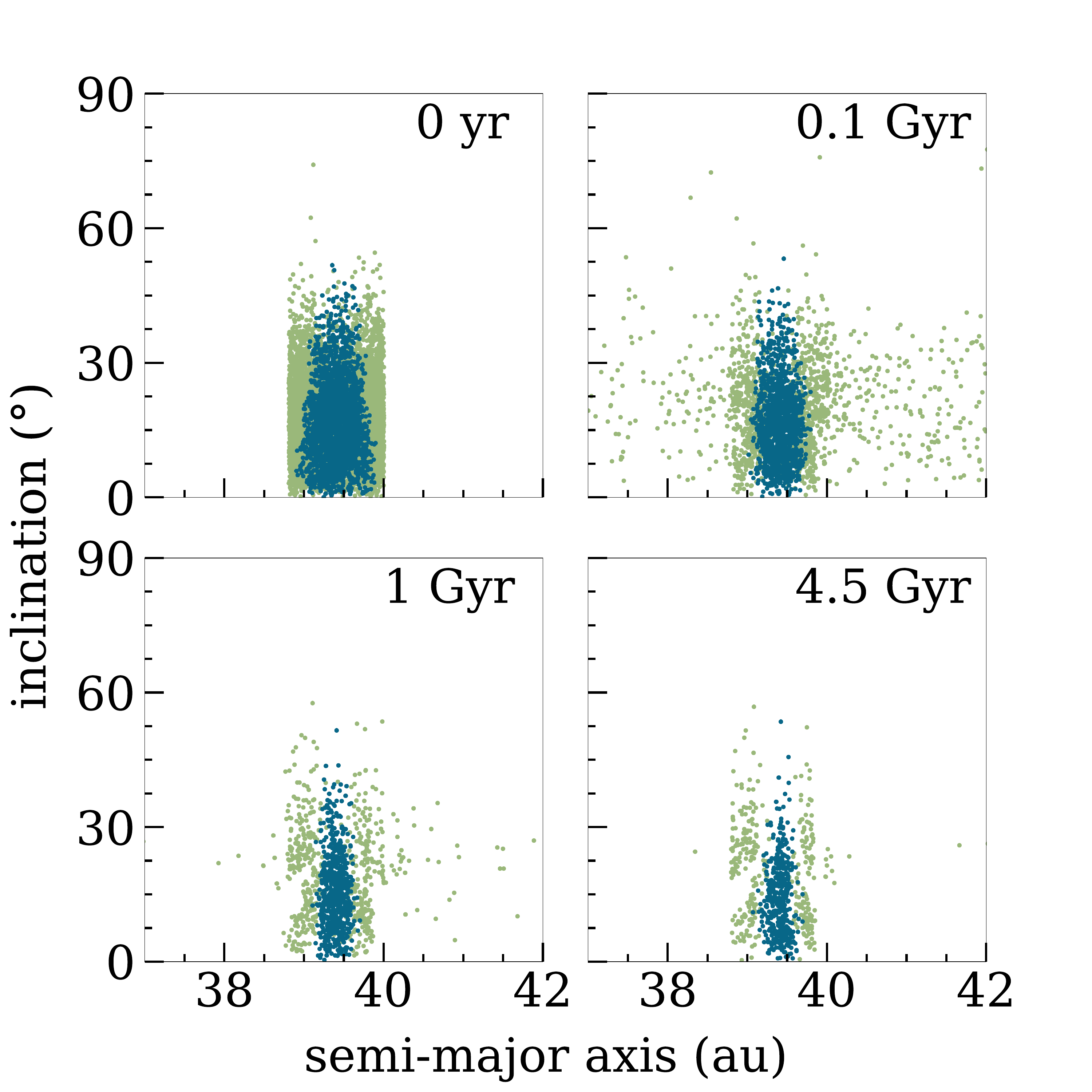}
        \caption{The inclination distribution of resonant particles does not evolve substantially over time.  
        We display inclinations for resonant (blue dots) and non-resonant (green dots) particles at four times for the same simulation shown in Figure \ref{fig:es}.  
        Initial inclinations are drawn from a modified Gaussian distribution (see Table~\ref{initParams}). 
        The simulation produces gaps in the non-resonant population on either side of the 3:2 resonance which we believe to be a secular resonance (For more details see \citealt{1991Icar...93..316K}).  
        In preliminary simulations (not shown), we found that resonant objects with inclinations spanning from $i=$0-90$^\circ$ remain stable over 4.5 Gyr, indicating that stability sculpting does not appreciably alter the inclination distribution in the resonance.  
        We thus chose an initial inclination distribution width of $\sigma_i=14^\circ$ which is consistent with the observed population \citep{volk2016}.}
    \label{fig:incs}
    \end{figure}

The distribution of particles in semi-major axis/inclination space is displayed in Figure \ref{fig:incs}. As expected, particles at the edge of the resonance are shaved over time, but the distribution of inclinations remains similar. 
As in our pilot simulations, we find no substantial correlation between the inclination and the stability of the particles in the resonance. A more in-depth discussion on the Plutino inclination distribution can be found in \cite{Li:2014_1}, \cite{Li:2014_2}, and \cite{Gomes:2003}. 
The diagonal gaps apparent in the non-resonant particles on either side of the 3:2 MMR in Figure \ref{fig:incs} likely result from a secular resonance that destabilizes particles at particular inclinations, as detailed in  \citealt{1991Icar...93..316K}. 
   
\begin{figure*}
        \centering
        \includegraphics[scale = 0.29]{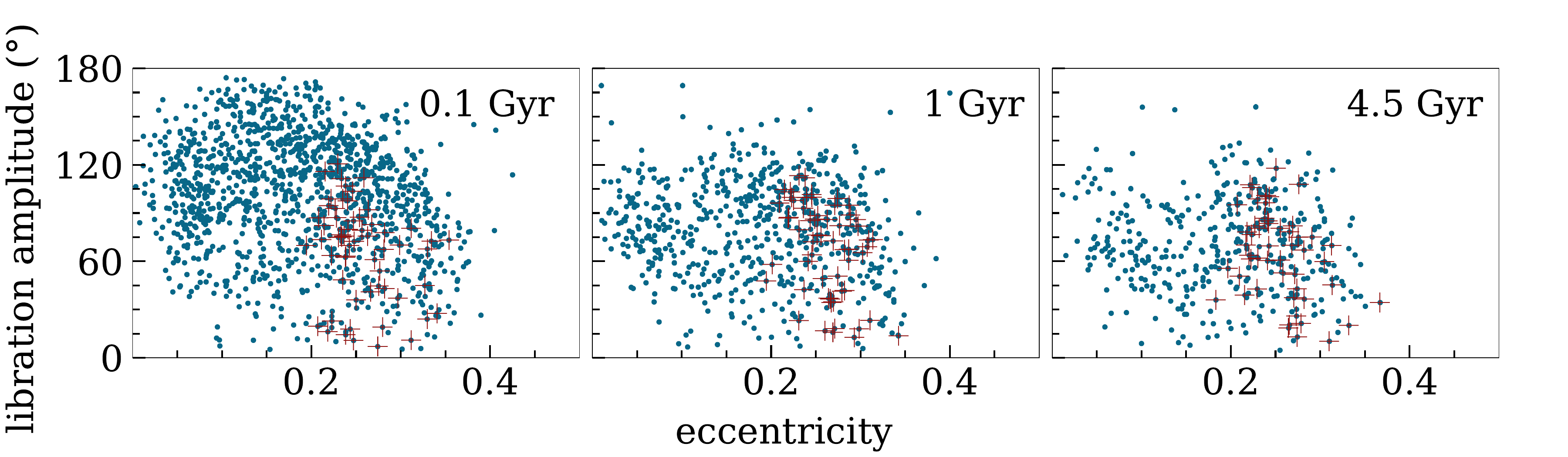}
        \caption{From 0.1 to 4.5 Gyr, the non-Kozai 3:2 resonant particles (blue dots) at higher libration amplitudes are not dynamically stable, whereas the Kozai 3:2 resonant particles (red crosses) remain stable at lower amplitudes with eccentricities $e\sim0.25$. 
        We see the largest decrease of high libration amplitude particles from 0.1 to 1 Gyr. 
        At 0.1 Gyr (left panel), 1 Gyr (middle panel), and 4.5 Gyr (right panel), 4.3\%, 8.7\%, and 11.5\% of the 3:2 resonant particles are also in the Kozai subresonance. 
        However, due to the relatively small sample size, the difference in the Kozai fraction from 1 Gyr to 4.5 Gyr is not statistically significant. 
        }
         \label{fig:kozailibe}
     \end{figure*}

Within the resonant population we are also interested in analyzing how the Kozai subresonance  evolves over time. 
At 0.1, 1, and 4.5 Gyr, the numbers of Kozai/resonant particles were 73/1698, 76/870, 64/556, respectively. 
While the number of resonant particles decreases significantly over time, the number of Kozai particles remains more constant.
The stable Kozai particles have eccentricities $e\approx0.25$ and their inclinations are distributed up to $i\sim45^\circ$.
Figure~\ref{fig:kozailibe} shows the libration amplitude vs. eccentricity for the Kozai and non-Kozai particles.

In general, resonant particles with higher libration amplitudes are preferentially lost over time.  
These objects are less stable because their resonant argument, $\phi$, deviates more from the central value $\pi$, allowing them to approach more closely to Neptune when they come to perihelion.  
As illustrated in Figure \ref{fig:kozailibe}, Kozai particles tend to have moderate-to-low libration amplitudes in the 3:2 MMR.
The lower 3:2 resonant libration amplitudes of Kozai objects likely contribute to their stability in addition to the libration of $\omega$ keeping the Kozai particles' perihelia locations away from the plane of the planets.

\section{OSSOS+ and Survey Simulator}\label{ossos}
    
To accurately compare our simulated 3:2 resonant population to the current observed population, we must account for  observational biases. 
Such biases are discussed extensively elsewhere \citep[see, e.g.,][]{Jones:2010,Lawler:2018ss}, but we review them briefly here. 
TNOs are detected by reflected sunlight, so detections are strongly biased against smaller objects and objects farther from the Sun; TNOs at perihelion are much more likely to be detected than those at aphelion, and large TNOs are more likely to be detected than small ones.
For objects in mean motion resonances, the resonant dynamics controls where objects come to perihelion relative to Neptune's position: KBOs in the 3:2 resonance come to perihelion preferentially $\pm90^\circ$ from Neptune.
This means that where observations occurred relative to Neptune will strongly influence the detectability of resonant objects (see \citealt{Gladman:2012} for a thorough discussion of this). 
Thus, accounting for observational biases in any given survey requires knowledge of the pointing history and well-determined limiting magnitudes for those pointings.

We compare our simulated 3:2 resonant population to the well-characterized sample of observed 3:2 resonant TNOs from several well-characterized surveys. 
We include 3:2 resonant objects from the A, E, L, and H observational blocks of the Outer Solar System Origins Survey (OSSOS) \citep{bannister2016,bannister2018}, as well as the 3:2 resonant objects from the Canada France Ecliptic Plane Survey (CFEPS) described by \cite{petit2011}, \cite{Gladman:2012};together these surveys comprise the OSSOS+ 3:2 resonance sample. 
The use of these detections to model TNO populations are described in, e.g., \cite{alexandersen2016} and \citealt{gutierrez:2019} among other works.
In this section we describe how we use the OSSOS+ survey simulator (described in Section~\ref{ss:surveysim}) to subject our simulated 3:2 resonant population to the same biases as the OSSOS+ observed 3:2 resonant population.
In Section~\ref{ss:rotation} we describe how we select and transform the orbital elements from our simulations to match them to a specific epoch near those of the OSSOS+ observations.
In Section~\ref{ss:hmag}, we describe how we then assign an $H_r$ magnitude to each set of orbital parameters (as all objects in our simulation are test particles, this part of the distribution is set based on prior studies).

\subsection{Survey Simulator}\label{ss:surveysim}

The OSSOS survey simulator software\footnote{https://github.com/OSSOS/SurveySimulator} is described in detail by \cite{petit2011} and \cite{Lawler:2018ss}. 
It is designed to take as input a TNO population model and output a list of simulated detections by subjecting that model to the observational biases of OSSOS and associated surveys (the OSSOS+ sample). 
These biases include the surveys' on-sky pointing histories, detection efficiency as a function of brightness and rate of motion, and the tracking/recovery efficiency for detected objects.

We feed the survey simulator a list of model TNOs, including their orbital elements at a specific epoch and their absolute magnitudes in r-band ($H_r$). 
These parameters fully describe the position and velocity of the model TNOs at a specific epoch from which the survey simulator can propagate them to all of the included observational epochs and, with $H_r$, determine their apparent magnitudes at these times. 
This full model of the 3:2 resonant population is run through the survey simulator to produce a large set of synthetic detections, i.e., what OSSOS+ would have observed if our model was representative of the true current 3:2 resonant population.

\subsection{Rotation}\label{ss:rotation}

The final locations of the giant planets in the simulations will not exactly match the locations of the planets at the epochs of the observations, so we must account for this when comparing to the observations.

This mismatch is not a problem during the orbital integrations because long-term dynamical stability depends on the average behavior of the planets over time rather than the specifics of the current epoch. 
However, we must correct for this difference when simulating detections because resonant objects are most detectable on-sky at specific longitudes relative to Neptune; it is thus necessary to rotate our simulation results to place the simulated Neptune near Neptune's current position to ensure that simulated resonant populations are oriented appropriately.  

To do this, we calculate the polar angle of Neptune's final location projected into the ecliptic plane, $\theta \equiv \tan^{-1}(y/x)$, where $x$ and $y$ are Cartesian coordinates in the ecliptic plane and $\hat x$ is the reference direction.  We then rotate every test particle's longitude of ascending node, $\Omega$, at the final timestep by the difference in Neptune's $\theta$ at the end of the integration and its $\theta$ from JPL Horizons at a reference epoch near the present.\footnote{We chose JD 2458970.5} 
This results in solid-body rotation of the entire system about the vertical ($z$) axis located at the barycenter of the solar system.

\begin{figure}
\centering
\includegraphics[scale = 0.3]{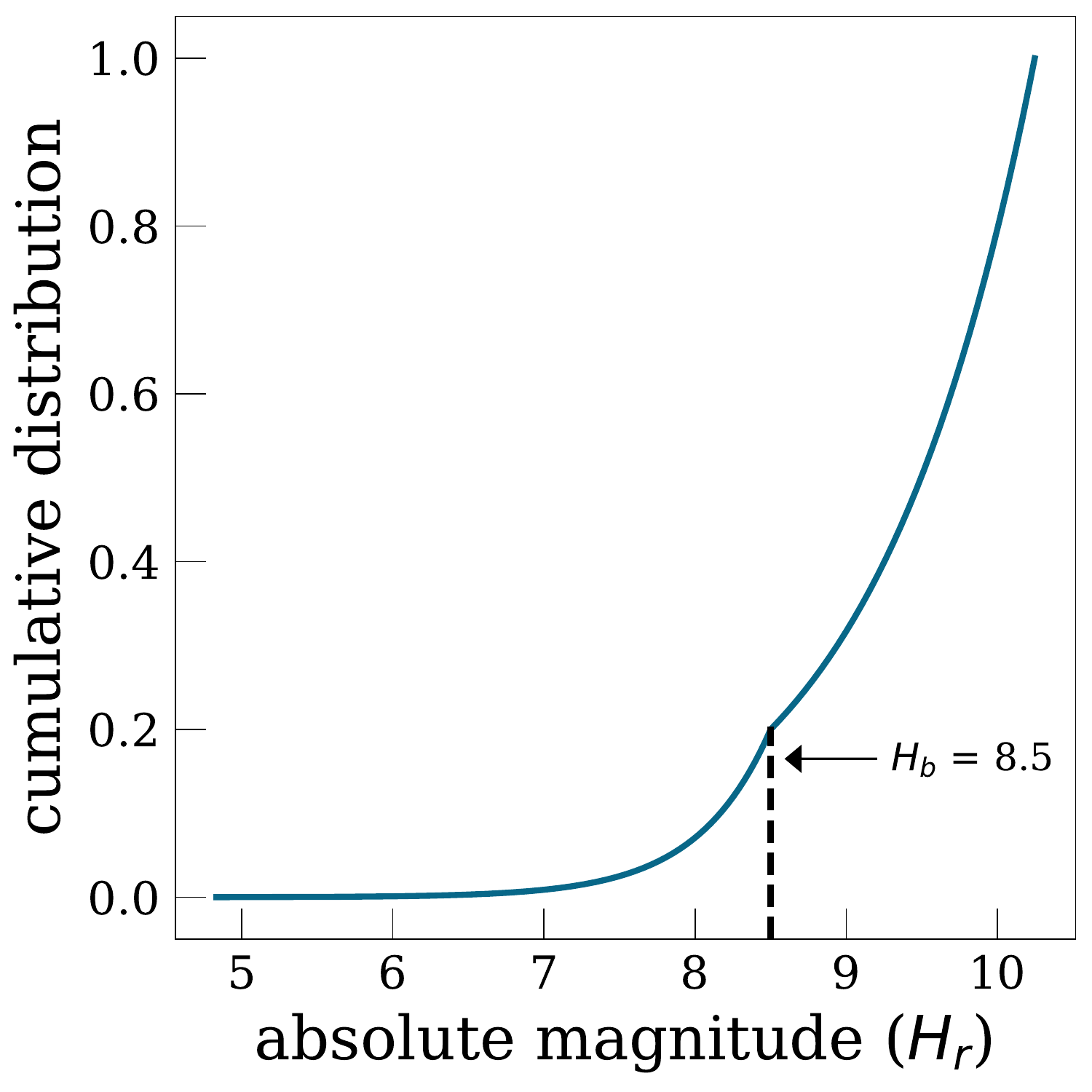}
    \caption{The absolute magnitude ($H_r$) distribution we adopt for our simulated 3:2 MMR population.  
    When passed to the survey simulator, each resonant particle in our model is assigned a value of $H_r$ randomly drawn from this distribution. 
    $H_r$ is modeled as a broken power law with a break magnitude ($H_b$) of $H_r = 8.5$. 
    For $H_r < H_b$, the distribution has an exponential slope of 0.9, referred to as the bright end slope ($\alpha_b$). 
    For $H_r > H_b$, the distribution has an exponential slope of 0.4, referred to as the faint end slope ($\alpha_f$). 
    In our distribution, the transition from $\alpha_f$ to $\alpha_f$ occurs at a break fraction of 0.2, meaning roughly 20\% of the objects will have $H_r < H_b$ and roughly 80\% of the objects will have $H_r > H_b$.}
    \label{hmag}
\end{figure}

\begin{figure*}
\begin{center}
\includegraphics[scale = 0.18]{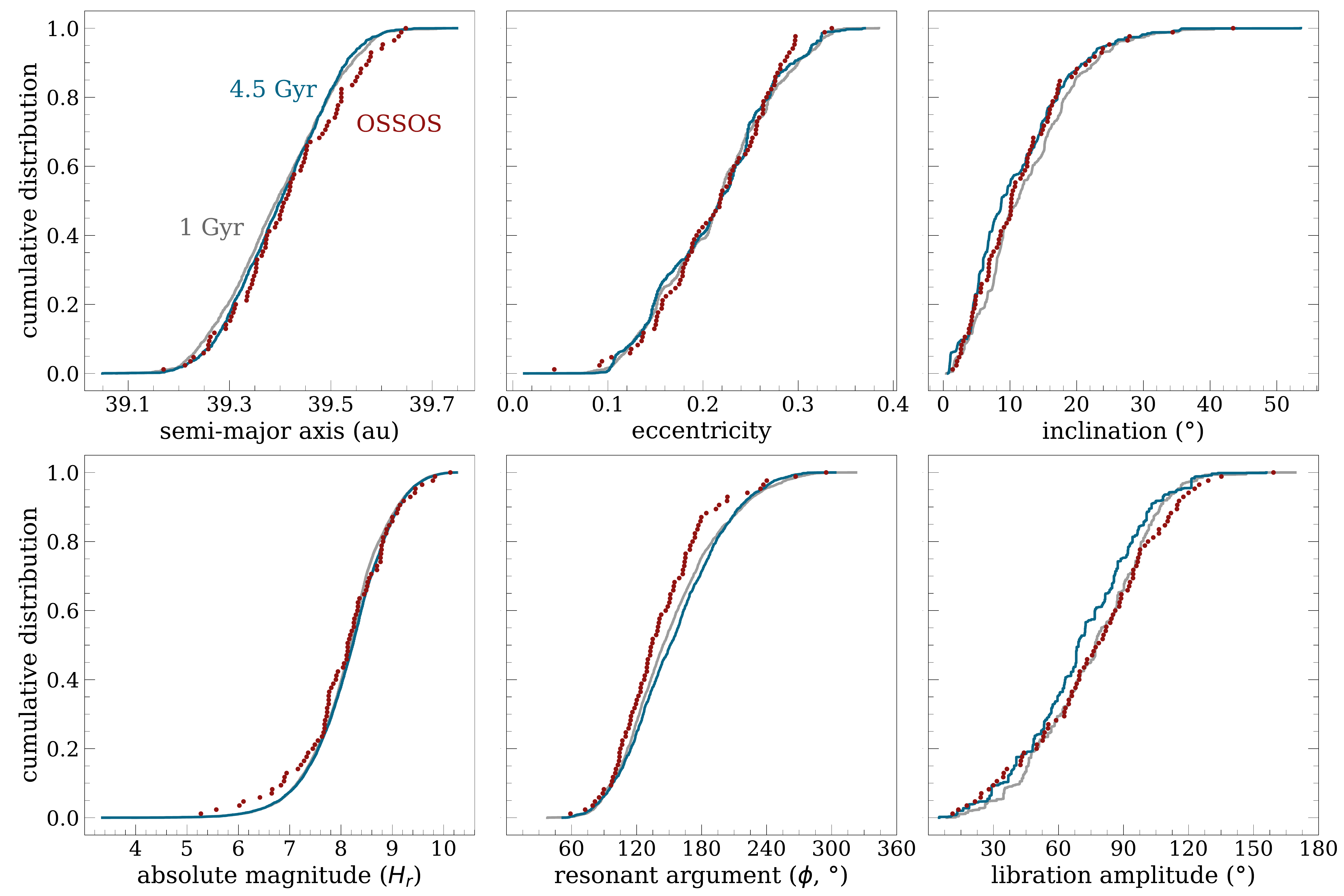}
\caption{When comparing our synthetic detections from the OSSOS Survey Simulator at both 1 Gyr (gray lines) and 4.5 Gyr (blue lines) against the OSSOS+ observed sample (red dots), we see a strong agreement between model and observations with the exception of our libration amplitude ($A_{\phi}$) distribution (bottom right panel). 
At 1 Gyr (grey, none of our six compared parameters are rejectable through the two-sample KS test or AD test: semi-major axis (top left), eccentricity (top middle), inclination (top right), absolute magnitude (bottom left), $\phi$ (bottom middle), libration amplitude (bottom right); see Table \ref{stats1}. 
However, at 4.5 Gyr we only see acceptable fits for semi-major axis, eccentricity, inclination, absolute magnitude, and $\phi$. Although the CDFs for $\phi$ appear to deviate significantly compared to the observed $\phi$ values at both 1 Gyr and 4.5 Gyr, we do not reject it since the Kuiper KS test accounts for the cyclical angular nature of $\phi$ and produces acceptable values at both time steps; see Table \ref{stats2}. 
When looking at libration amplitude, we see that despite the 1 Gyr model deviating from observations at the tails of the distributions, there is a close match in the middle of the distribution which allows it to produce acceptable statistics. 
However the libration amplitude distribution at 4.5 Gyr only matches well at low libration amplitudes ($\lesssim 50$) which results in rejectable statistics.}
\label{fig:cdfs1}
\end{center}
\end{figure*}

\subsection{Cloning, Color distribution, and H-magnitudes}\label{ss:hmag}

The number of 3:2 resonant particles in our simulation at any single snapshot in time is far fewer than the number needed for the survey simulator to produce a large enough sample of synthetic detections to robustly compare with OSSOS+ data.  
After 4.5 Gyr, 556 particles remain in the 3:2 resonance in our simulation.  
While this number is sufficient to map the phase space of the resonance well if all particles are considered, at any given snapshot in time, many particles will be un-observable.
A typical 3:2 resonant object is small and only visible near the pericenter of its orbit---near apocenter, it is too distant from the Sun and thus too faint to be seen. 
We thus ``clone" each test particle to sample a large range of phases along its orbit. 

We take the orbital parameters of each particle at each timestep in the short $10^5$-year integration (started at either 1 or 4.5 Gyr, depending on the comparison being made) and treat it as a new particle, essentially ``cloning" the actual test particle into 1000 pseudo-particles. 
Having 1000 clones of each resonant particle ensures that we have enough simulated detections from the OSSOS Survey Simulator to have reliable statistics when we compare our models to the OSSOS+ observations.
    
To forward-bias our models with the OSSOS Survey Simulator, several things are required: positional information for each object in the model, an $H_r$ magnitude for each object in the model, a color distribution, and an epoch. 
Our \textsc{rebound} simulations give us the positional information we need in the form of the six orbital elements: $a$, $e$, $i$, $\Omega$, $\omega$, and $M$. 
We add an $H_r$ magnitude to each object, a color distribution (to account for the fact that some of the OSSOS+ 3:2 objects were discovered in different filters), and an epoch to the output of the simulation before running the particles through the OSSOS Survey Simulator.
    
For the $H_r$ magnitude, we use a broken power law size distribution derived from a modified version of Equation 4 from \cite{volk2016}.  
A broken power law in size corresponds to two exponentials in absolute magnitude $H_r$ affixed at a specified break magnitude. 
Our choice of distribution is displayed in Figure~\ref{hmag}. 
The distribution is normalized by specifying the cumulative fraction of objects over the full modeled $H_r$ range that are below the break magnitude.  
We choose a bright-end slope of 0.9 based on previous modeling of the OSSOS 3:2 resonant population \citep{volk2016}. 
We tested a range of values drawn from literature constraints \citep[e.g.][]{ Shankman:2013, Fraser2014, alexandersen2016,Lawler:2018} for the break magnitude and faint-end slope. We choose a break magnitude of $H_r=8.5$, a break fraction of 0.2, and a faint end slope of 0.4, which provide a good match for the observed eccentricity distribution (see Figure~\ref{fig:cdfs1} in Section~\ref{discussion}.) 
Each object in the simulation output is attributed a random $H_r$ sampled from this distribution.

For the color distribution, we use the same approach as in the CFEPS L7 model \citep{petit2011},  with a few modifications. 
The color distribution used by \cite{petit2011} works by assigning the $H_r$ magnitude as the magnitude in a specified color band to be used as a reference. 
For their distribution, \cite{petit2011} chose the g-band to be the color used when specifying the $H_r$ magnitude. 
The magnitudes in other bands were calculated from shifting up or down from the g-band. 
We use this same distribution for our models, but we use the r-band as the reference band since the OSSOS observations were done in the r-band and dominate the sample we are comparing to \citep{bannister2018}. 
We define the g-r color to be 0.65 based on recent observations \citep{schwamb2019}. 
We do not change any of the other conversions from \cite{petit2011}, as the g-band and r-band were the only two filters used for discovery in the OSSOS+ ensemble \citep{petit2011,alexandersen2016,bannister2018}.

\section{Statistical Comparisons}\label{discussion}

To test the rejectability of our models, we compare our forward biased models to the OSSOS+ detections by performing the two sample Kolmogorov-Smirnov (KS) test and Anderson-Darling (AD) on the distributions of $a$, $e$, $i$, $H_r$, $\phi$, and $A_{\phi}$. 
We also utilize the Kuiper variant of the KS test specifically when looking at $\phi$, it being a better test to use when comparing distributions of cyclical angular quantities. 
The null hypothesis, $H_0$, of each test is the same: the two distributions being compared could have been drawn from the same parent distribution.  
Though the KS, AD, and Kuiper-KS tests are simple 1D statistics that can only test for rejectability, not goodness of fit, they are frequently used for comparisons of populations in the trans-Neptunian region because the complicated phase space of orbits renders more detailed statistical analysis computationally prohibitive unless one is restricted to a small region of phase space \citep[see, e.g.,][Appendix A]{volk2016}. While we compare the distributions of the six mentioned values, we are not aiming to explain the origin of the inclination or magnitude distributions. We assume the inclination distribution is formed before Neptune reaches its final semi-major axis of $a=30.1$~au and the magnitude distribution is set by formation processes not discussed in this paper.

We begin by calculating a test statistic unique to the each of the three tests. 
The KS test statistic, $D_{KS}$, is defined to be the maximum vertical distance between the cumulative distribution functions (CDFs) of the two distributions being compared; for the Kuiper variant\footnote{Based on NIST handbook: \url{https://www.itl.nist.gov/div898/handbook/eda/section3/eda35e.htm}}, $D_{Kuiper}$ is defined to be the sum of the maximum and minimum vertical distances between the CDFs. 
The AD test statistic, $D_{AD}$ is similar to $D_{KS}$, but gives more weight to differences towards the tails of the distribution, while the KS test is dominated by differences in the middle of the distribution (because the CDFs for each distribution are forced to be 0 and 1 at either end of the distribution). 
For both $D_{KS}$ and $D_{AD}$, we use the functions built into the $\emph{SciPy's}$ Python package to calculate the test statistics.

After calculating the test statistic, we use a Monte Carlo sampling method to calculate a p-value for the result; our p-value is defined as the fraction of N synthetic test statistics generated by comparing the model to itself that were greater than the calculated test statistic when comparing the model to the observations.  
The rejectability of $H_0$ is $1-p$. 
We place a 95\% confidence limit on our p-values, meaning we reject $H_0$ if $p<0.05$. 

There are 85 observed 3:2 resonant objects in the OSSOS+ survey. As such, we randomly select 85 objects from our forward biased 3:2 resonant model and calculate the test statistic between this random sample and the full forward biased 3:2 resonant model. This process is repeated N times to yield N test statistics.
To obtain consistent p-values using this method, we find that at least 100,000 random draws are needed.

\subsection{Our model vs OSSOS+}\label{stats}

Recalling that the null hypothesis we are testing for is that the OSSOS+ sample and our forward-biased 3:2 MMR model could have come from the same distribution, we perform the analysis described above for the parameters  $a$, $e$, $i$, $H_r$, $\phi$, and $A_{\phi}$, at both 1 Gyr and 4.5 Gyrs (see Figure \ref{fig:cdfs1}).
When we feed our full model of the 3:2 population through the survey simulator, we find that we cannot match the OSSOS eccentricity distribution because too many low-eccentricity objects are detected. We therefore consider the likely possibility that objects were not scattered from pericenter distances extending all the way out to the current location of the resonance at 40au.

To investigate the potential that the 3:2 resonance was populated with particles scattered outward from a more limited rage of initial heliocentric distances, we apply a cut in our initial test particle distribution to remove particles with initial pericenter distances larger than values ranging from 33-38 au in 1 au increments.
These six resulting models (which are subsets of our total simulation data) are fed through the Survey Simulator, and we find good agreement with the observed eccentricity distribution for pericenter cuts between 35 and 37au, while cuts at $q= 33$, 38, and 39au are rejected by the KS-test and cuts at $q= 33$, 34, 38, and 39au are rejected by the AD test. The best fit arises when objects having initial pericenters greater than 36 au are removed. All further results presented here include a 36au pericenter cut, corresponding to an initial scattering region ending at 36 au. 

With this pericenter cut, at 1 Gyr, we do not reject the null hypothesis for any parameters, whereas at 4.5 Gyr, $A_\phi$ and $\phi$ produce rejectable p-values below 0.05. The angle $\phi$ is cyclical however, so we perform a Kuiper KS test which is designed for cyclic angles. The p-value for this test is above 0.05, so we conclude $\phi$ falls in line with the null hypothesis (see Tables ~\ref{stats1} and~\ref{stats2}). 

\begin{table}
\caption{Statistics results for OSSOS+ detections vs simulated detections at 1 Gyr. 
Results at 1 Gyr are non-rejectable because all p-values lie above 0.05, indicating that we are within 95\% confidence for all p-values. }
\vskip0.1in
\begin{tabular}{lllllll}
\hline
 & $D_{KS}$ & $p_{KS}$ & $D_{AD}$ & $p_{AD}$ & $D_{Kuiper}$ & $p_{Kuiper}$\\
\hline
$a$ & 0.107 & 0.252 & 1.584 & 0.07 & - & -\\
\hline
$e$ & 0.087 & 0.463 & -0.258 & 0.473 & - & -\\
\hline
$i$ & 0.099 & 0.352 & -0.079 & 0.389 & - & -\\
\hline
$H_r$ & 0.081 & 0.577 & -0.174 & 0.428 & - & -\\
\hline
$\phi$ & 0.118 & 0.172 & 1.29 & 0.09 & 0.136 & 0.415\\
\hline
$A_{\phi}$ & 0.084 & 0.519 & 0.299  & 0.25 & - &-\\
\hline
\end{tabular}
\label{stats1}
\end{table}

\begin{table}
\caption{Statistics results for OSSOS+ detections vs simulated detections at 4.5 Gyr. 
Results at 4.5 Gyr are non-rejectable for all quantities except $A_{\phi}$ as the p-values for both tests performed on it fall below our limit of 0.05. 
Although the p-values for the KS test and AD test fall below 0.05 for $\phi$, we do not reject it because the p-value for the Kuiper variant of the KS test lies above our limit of 0.05.}
\vskip0.1in
\begin{tabular}{|l|l|l|l|l|l|l|}
\hline
 & $D_{KS}$ & $p_{KS}$ & $D_{AD}$ & $p_{AD}$ & $D_{Kuiper}$ & $p_{Kuiper}$\\
\hline
$a$ & 0.108 & 0.241 & 1.525 & 0.073 & - & -\\
\hline
$e$ & 0.09 & 0.429 & -0.051 & 0.372 & - & -\\
\hline
$i$ & 0.114 & 0.197 & 0.222 & 0.276 & - & -\\
\hline
$H_r$ & 0.087 & 0.479 & -0.219 & 0.448 & - & -\\
\hline
$\phi$ & 0.159 & 0.023 & 3.855 & 0.009 & 0.176 & 0.08\\
\hline
$A_{\phi}$ & 0.143 & 0.046 & 3.831 & 0.009 & - & -\\
\hline
\end{tabular}
\label{stats2}
\end{table}

 \begin{figure}
    \begin{center}
    \includegraphics[scale = 0.22]{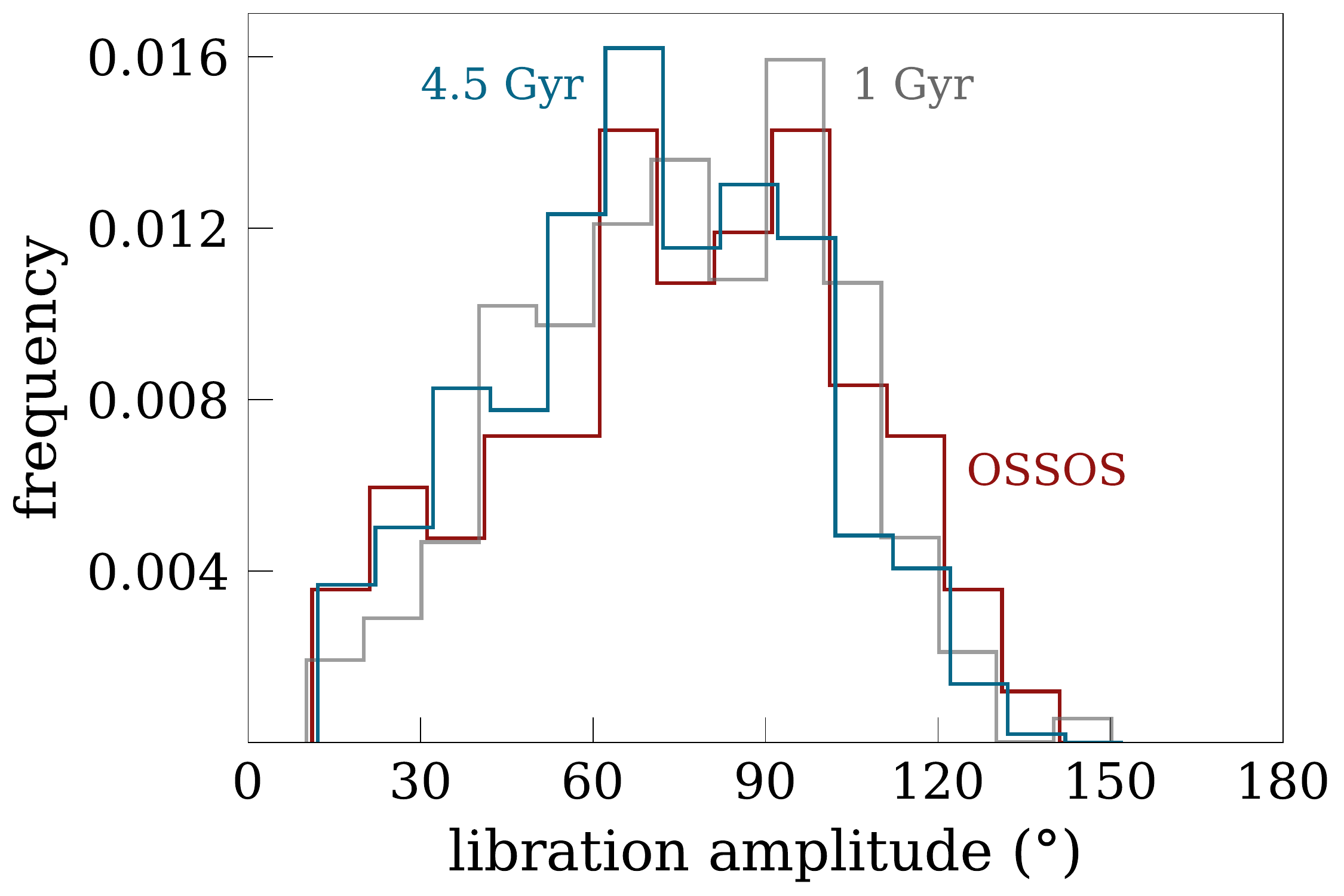}
    \caption{An alternative view of the bottom right panel from Figure \ref{fig:cdfs1} which further highlights the differences in libration amplitude ($A_{\phi}$) distributions between the simulated detections from our model and the OSSOS observations. 
    The 1 Gyr (gray), 4.5 Gyr (blue), and OSSOS (red) samples contain 10632, 7927, and 85 objects, respectively. 
    We see that the 4.5 Gyr model distribution is weighted slightly more toward low libration amplitudes compared to the 1 Gyr model distribution (as expected due to loss of high-amplitude 3:2 objects over time) and compared to the OSSOS sample.
    Note that we offset each of the histogram curves by $1^\circ$ for clarity in distinguishing them.}
    \label{fig:libamp1}
    \end{center}
   \end{figure}

\begin{figure}
    \centering
    \includegraphics[scale=0.8]{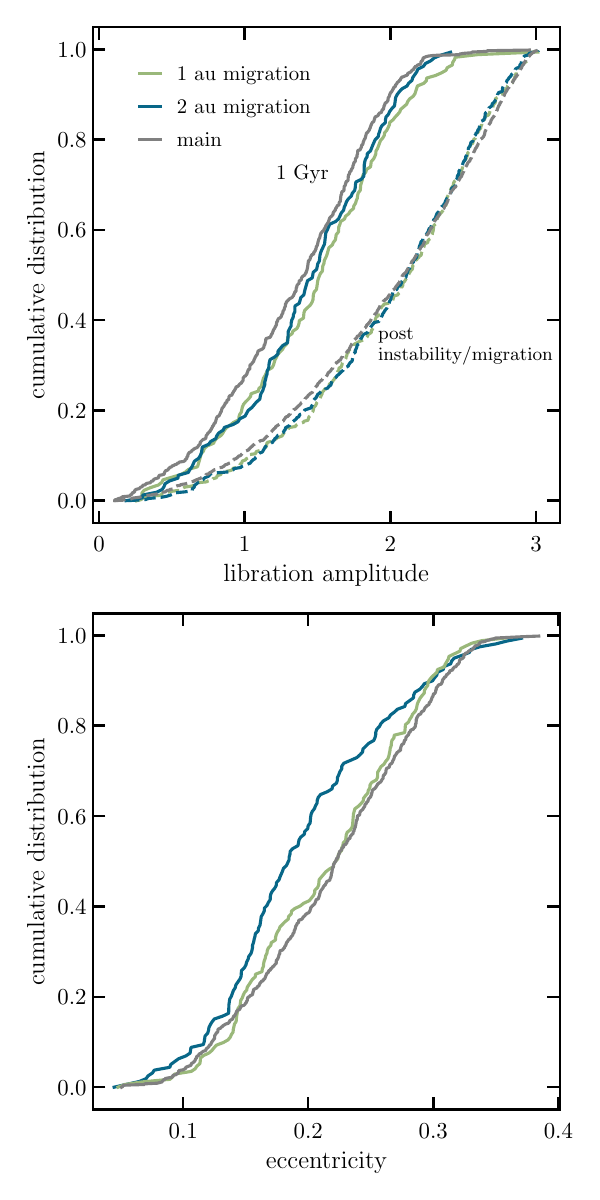}
    \caption{
    The 3:2 resonant population's libration amplitude distribution (top panel) after a 10 Myr migration of 2 au or less (dashed blue and green lines) does not differ from the "post instability" initial resonant population (grey dashed line). At 1 billion years, the libration amplitude distribution between the three models also doesn't differ significantly (solid lines). The eccentricity distribution does not differ for migration distances of 1 au or less at 1 billion years (bottom panel). Intrinsic distributions resulting from the simulations are shown (unlike in Figure \ref{fig:cdfs1}, these populations were not passed through the OSSOS survey simulator).  We find that a brief epoch of smooth migration does not materially change our results.}
    \label{fig:smoothmigration}
\end{figure}
   
\subsubsection{Libration Amplitude, $A_\phi$}
\label{libamp}
An alternate view of the $A_{\phi}$ distributions is shown in Figure~\ref{fig:libamp1} to show the discrepancy between the synthetically detected objects from the simulation and the OSSOS+ observations in more detail. Alternative pericenter cuts did not improve agreement.

The discrepancy at the current solar system age of 4.5 Gyr is significant but modest. Within the context of the model considered here, two possibilities for resolving it immediately present themselves. First, transient sticking \citep[e.g.,][]{Lykawka:2007,yu2018} adds a pseudo-stable population of particles to the resonance at preferentially high libration amplitudes.  OSSOS objects are identified with million-year integrations and their longer-term resonance stability time is not currently available.  The objects in our sample are stable over billion year timescales. In other words, the observations should contain high-libration-amplitude transient objects which our model does not.
Whether the transient sticking population adds sufficiently many high-libration-amplitude objects to resolve the discrepancy merits future work.  We consider this possibility promising. 

Alternatively, planetary upheaval models require that Neptune's eccentricity ultimately be damped to its current low value.  This damping is thought to result from dynamical friction with planetesimals, a process which also results in smooth migration. While dynamical friction in a symmetric sea of particles normally results in the planet's inward migration from angular momentum transfer, in the case of the outer solar system, the ice giants migrate outward. This is due to an asymmetry between the number of planetesimals from which Neptune takes angular momentum and the number that give angular momentum to Neptune. This global asymmetry results from the presence of the other giant planets (see \citealt{fernandez:1984} and \citealt{Tsiganis:2005} for more details.)

Since smooth migration pushes objects more deeply into resonance, such a late-stage epoch of migration has the potential to modify the distribution found here, either in the direction of better or worse agreement.  We investigate the impact of post-upheaval smooth migration on libration amplitudes with 4 independent smooth migration simulations including the giant planets and 8000 test particles. In the simulations, Jupiter, Saturn, and Uranus begin at their current locations and Neptune at $\sim$ 29, 28.5, 28, and 27 au respectively. Neptune migrates for 10 million years up to its barycenter value of $\sim$ 30.06 au for all simulations, and we continue to integrate up to 1 billion years to compare with the 1 billion year simulation in this paper. The test particles are initialized with similar distributions as those in our main simulation, but with a broader range of semi-major axes.  For each value of Neptune’s initial semi major axis, we fill the phase space with test particles from the interior edge of the 3:2 resonance before migration to the exterior edge  after migration.

We find that the libration amplitude distribution for 3:2 resonant objects does not differ from our non-migrating simulation when the migration distance is $\lesssim$2 au and the eccentricity distribution does not differ for migration distances $\lesssim$1 au, as illustrated in Figure \ref{fig:smoothmigration}. Thus a brief epoch of smooth migration neither improves nor worsens the match between our model and the OSSOS libration amplitude distribution. We note that exploration of larger migration distances would necessitate adjusting our pericenter cut, running separate 4.5 Gyr simulations for each migration scenario, and running these through the OSSOS survey simulator, which we reserve for future work.

   \begin{figure}
    \centering
    \includegraphics[scale = 0.55]{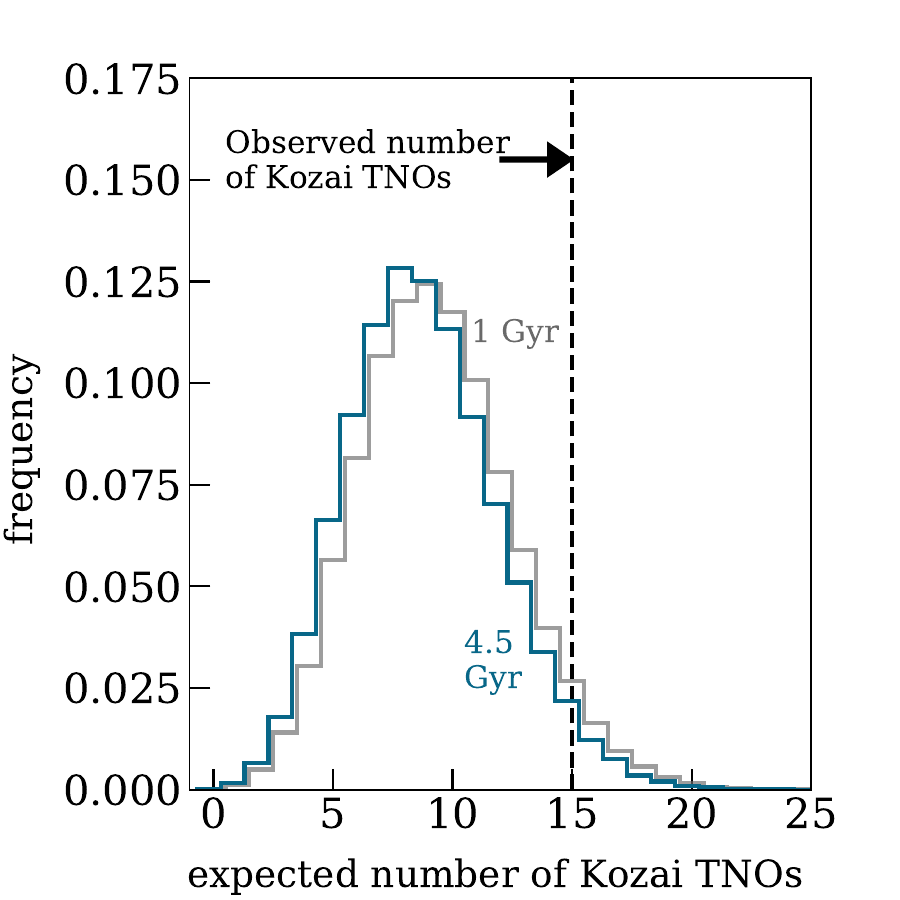}
        \caption{Our model produces significantly fewer detected Kozai librators in the 3:2 MMR than observed in the OSSOS+ sample.
        We use a Monte Carlo sampling method (see Section \ref{sss:kz}) to generate the expected distribution of how many 3:2 TNOs in the Kozai resonance would be included in a total sample of 85 detected 3:2 objects. There are 15 observed Kozai librators (see Section~\ref{sss:kz}) in the OSSOS+ sample (using the same $\omega$ libration cut as in our simulated sample) which is larger than the expected number of synthetic detections at both 4.5 Gyr (blue) and 1 Gyr (gray). 
        The histogram for 4.5 Gyr is shifted by 0.2 to the left for clarity.}
        \label{fig:kz_hist}
    \end{figure}

\begin{figure}
    \begin{center}
    \includegraphics[scale = 0.3]{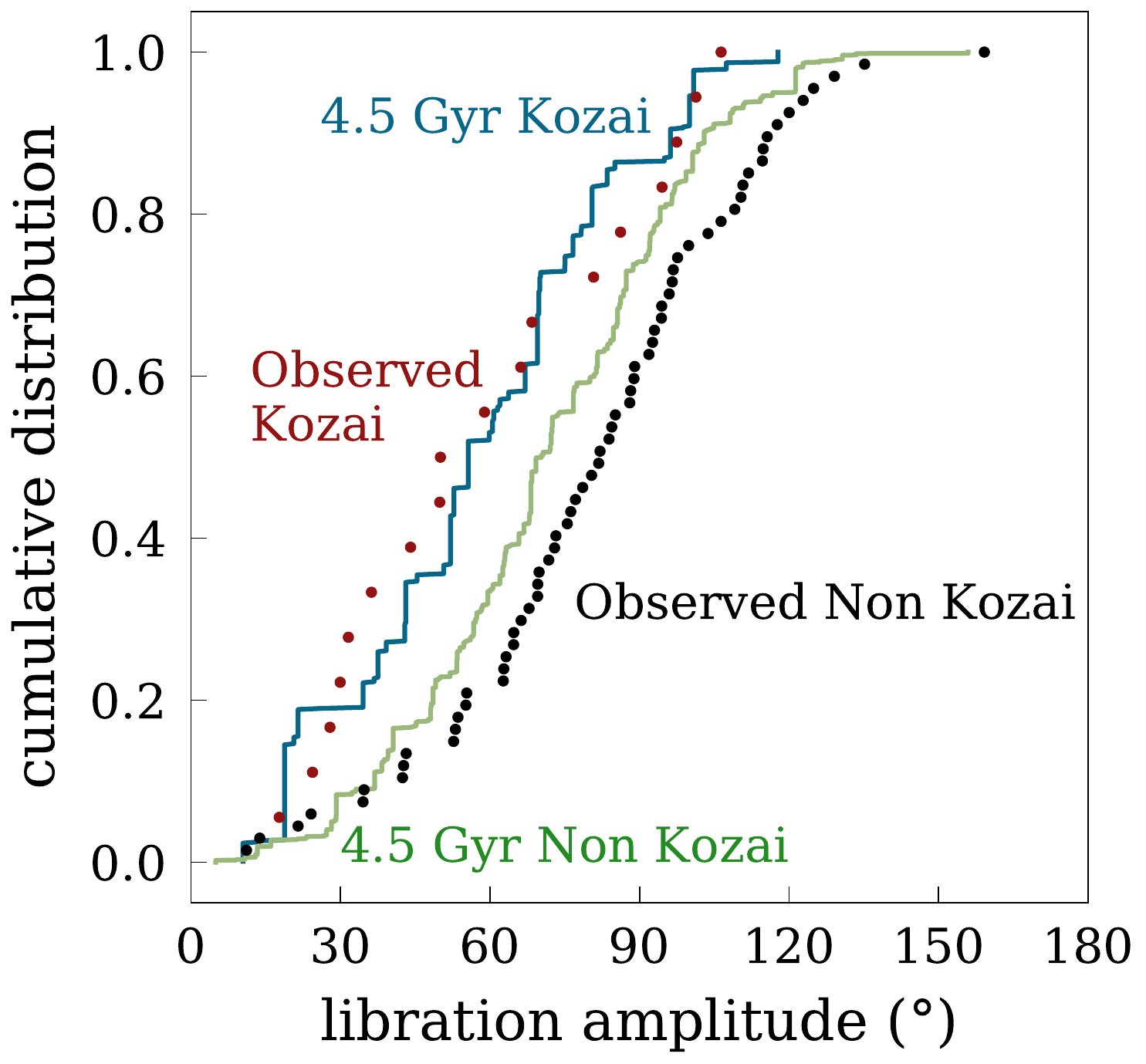}
    \caption{The detected Kozai and non-Kozai populations have significantly different libration amplitude distributions in both the simulated and OSSOS+ detections. 
    In particular, the simulated detected Kozai objects in our model at 4.5 Gyr (blue line) and the observed OSSOS+ Kozai objects (red dots) tend towards lower libration amplitudes and follow a similar distribution. 
    The simulated detected non-Kozai objects in our model (green line) and the observed OSSOS non-Kozai objects (black dots) both tend towards higher libration amplitudes, but they do not match each other as well as their Kozai counterparts.}
    \label{fig:cdfs2}
    \end{center}
   \end{figure}

\subsubsection{Kozai population}\label{sss:kz}

We compare the expected Kozai subpopulation of the 3:2 resonance from our simulations to the observations to further examine the accuracy of our model. 
We use a Monte Carlo sampling method for this comparison.
Taking the 3:2 resonant particles with initial pericenters below 36 au from our model that are detected by the survey simulator, we randomly draw samples of 85 3:2 objects and then count how many of those 85 simulated detections are of Kozai particles. 
We repeat this process $10^5$ times for both the 1 and 4.5 Gyr simulation snapshots to produce the distribution of expected observed Kozai particles shown in Figure~\ref{fig:kz_hist}.

Interestingly, the Kozai fraction in our raw simulation (i.e. without going through the survey simulator) increased from 11.1\% of 3:2 resonant objects at 1 Gyr to 14\% at 4.5 Gyr, but Figure~\ref{fig:kz_hist} shows that the expected number of \textit{detected} Kozai objects is nearly identical at both simulation times. 
While this apparent contradiction could possibly be related to the very complex observational biases in the Kozai population (see, e.g., \citealt{Lawler:2013}), it is also possible that it is due to the relatively small number statistics of Kozai objects in our simulations; using simple Poisson error estimates, the Kozai fractions in our simulations at 1 and 4.5 Gyr are marginally consistent with each other (though we note that because Kozai 3:2 resonant particles are more stable than non-Kozai , an increase in Kozai fraction over time is expected!).

As mentioned in Section~\ref{initParams}, we identify the Kozai objects in the simulation by checking if their $\omega$ librates between 5$^{\circ}$ and 175$^{\circ}$ or 185$^{\circ}$ and 355$^{\circ}$. 
In the OSSOS dataset considered here, there are 18 3:2 objects that are in the Kozai subresonance.
However, we find that if we restrict the libration of the observed objects to the same ranges, our check for Kozai fails to catch 3 real observed objects with libration centers other than $90^\circ$ and $270^\circ$ (these are classified as Kozai largely based on visual examination of their orbital histories). 
We thus compare our simulation results to the 15 real observed Kozai 3:2 objects that librate in the same way as our simulated ones.
Figure~\ref{fig:kz_hist} shows that at both simulation snapshots, the number of simulated observed Kozai 3:2 objects is significantly smaller than the number observed by OSSOS. Out of 100000 total draws, 95.1\% of draws contained $<15$ Kozai objects.

To check whether the rejectability of the model's predicted Kozai fraction and the rejectability of the predicted $A_{\phi}$ distribution are potentially related, we examine the libration amplitude distribution of the Kozai and non-Kozai 3:2 particles separately; this is shown in Figure \ref{fig:cdfs2}. 
Both the real and synthetic detected Kozai 3:2 populations are weighted toward smaller libration amplitudes (consistent with what we saw in our intrinsic model population; see Figure~\ref{fig:kozailibe}).
Because the discrepancy in Figure \ref{fig:cdfs2} arises from the non-Kozai objects, we confirm two unrelated discrepancies: an under population of Kozai objects and underpopulation of mid-high libration amplitudes.

Upon running smooth migration simulations, introduced in Section \ref{libamp}, we found that all simulations had twice as many or more objects in Kozai resonance than before migration. When comparing the raw smooth migration simulation and main simulation discussed in this paper (i.e. without running them through the OSSOS survey simulator), we found that at 1 billion years, the 2 au smooth migration model had 13 \% objects in kozai whereas the intrinsic simulation had 11 \% in kozai at 1 billion years. While 13 \% is higher than 11 \%, we do not believe it’s significant enough to confidently say smooth migration will increase kozai objects significantly. We will explore this more rigorously in future work.

\begin{figure*}
    \centering
    \includegraphics[scale = 0.25]{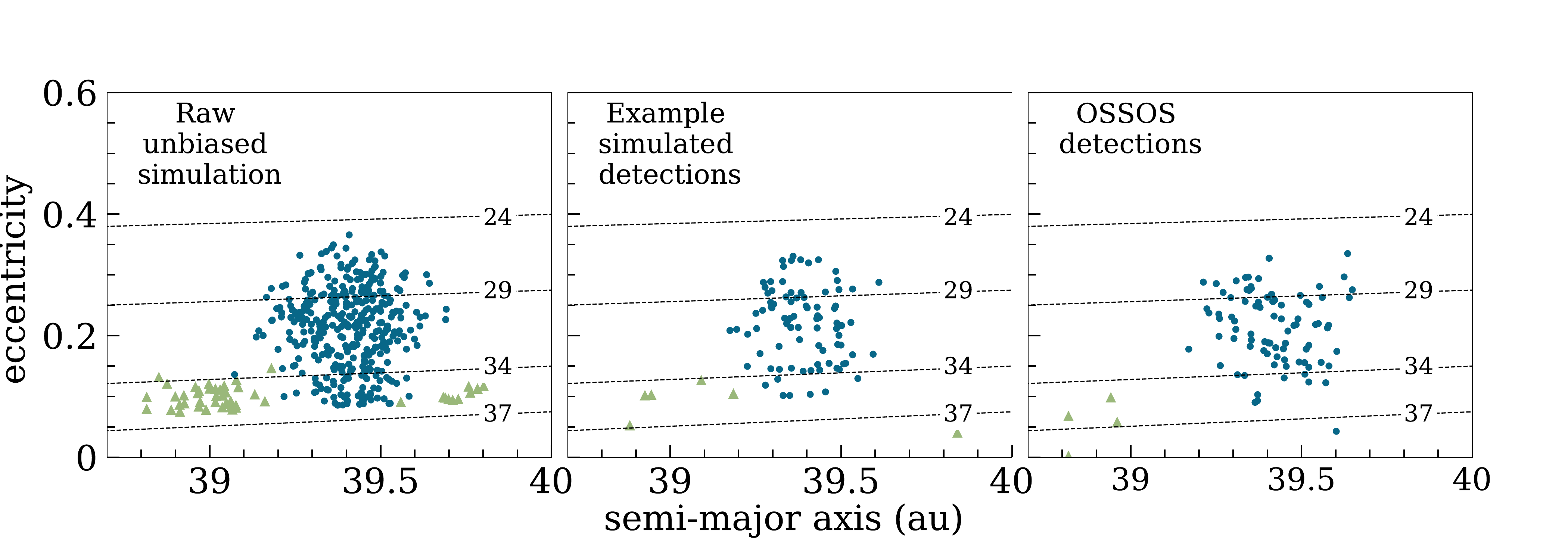}
        \caption{When we pass the raw simulation data (left panel) through the survey simulator and sample the resulting simulated detections (middle panel) they reproduce the distribution of real OSSOS+ detections (right panel) reasonably well. 
        The middle panel shows a single iteration of the Monte Carlo sampling method used to compare the number of simulated stable classicals (green triangles) for every 85 3:2 TNO's that would be synthetically detected at 4.5 Gyr in the region $a = 38.81-40$ au with a pericenter cut at 36 au. 
        The expected distribution of detected stable classicals from our model in $a-e$ space is similar to those detected by OSSOS+ in the same $a$ range.
        The similarity between the observed classicals and synthetically detected classicals illustrates the validity of the $q=36$ au cut.}
    \label{fig:cc_sampling_ex}
    \end{figure*}

    \begin{figure}
    \centering
    \includegraphics[scale = 0.55]{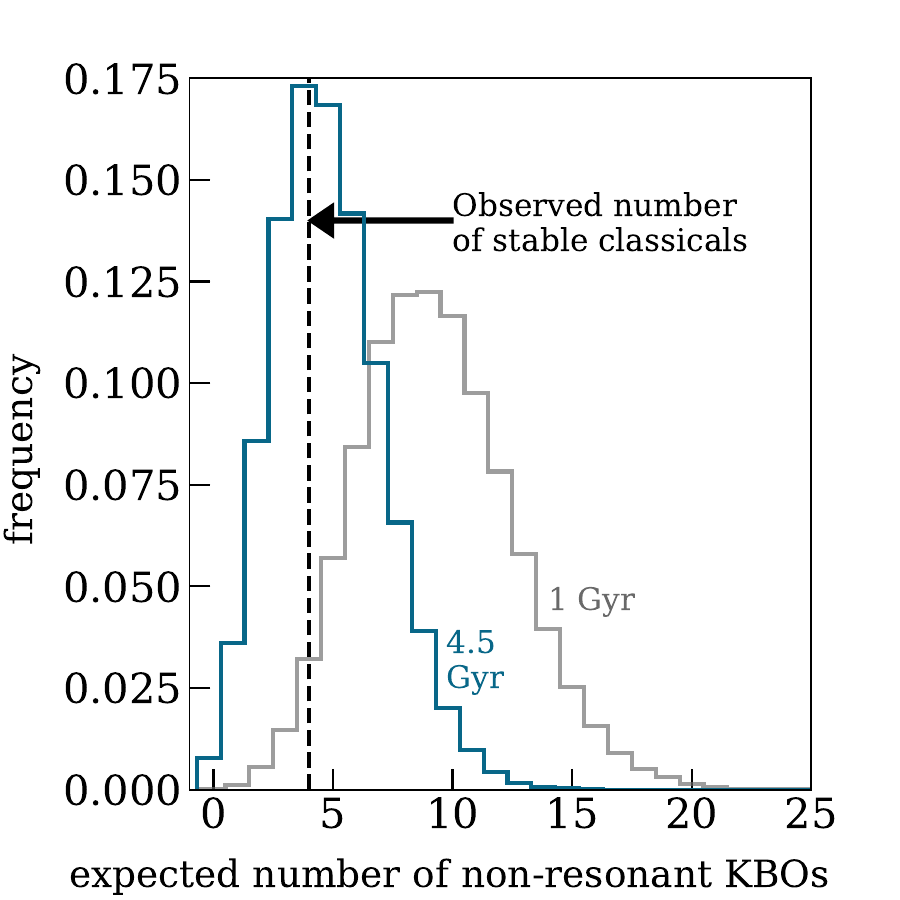}
        \caption{The population of stable nonresonant objects from our simulation that are synthetically detected by the survey simulator at 4.5 Gyr is consistent with the observed number of classicals in the region $a = 38.81-40$ au. 
        We apply the same pericenter cut at 36 au as in Figure \ref{fig:cdfs1} to the simulated data and use a Monte Carlo sampling method (see Section \ref{sss:cls}) to find how many classicals would be detected in our simulation for every 85 3:2 objects drawn. 
        There are 4 observed OSSOS+ classicals in the aforementioned range which is consistent with the 4.5 Gyr curve (blue). 
        The same sampling method was used at the 1 Gyr snapshot (gray) and is shown for reference. 
        The histogram for 4.5 Gyr is shifted by 0.2 to the right for clarity. 
        }
    \label{fig:c_sampling}
    \end{figure}

\subsubsection{Classical population}\label{sss:cls}

Figures~\ref{fig:es} and~\ref{fig:incs} show that some of the non-resonant test particles in the vicinity of the 3:2 survive our 4.5 Gyr simulations. 
This provides an additional observational test for the perihelion distance cut used to best reproduce the observed 3:2 population. 
Using this same perihelion distance cut at 36 au, we can examine how many classical (non-resonant) objects OSSOS+ should have observed in the region immediately surrounding the 3:2 resonance if the initial phase space was filled as in our model.

We compare the expected number of observed stable non-resonant TNOs from the simulation (see Figure~\ref{fig:es}) at 1 Gyr and 4.5 Gyr with the observed number in the OSSOS+ sample by considering the sample of all test particles (resonant and non-resonant) in the restricted $a$ range of 38.81-40 au with initial $q>36$ au (the $q$ cut determined in Section \ref{ss:hmag}).
We pass all of these test particles, resonant and non-resonant, through the survey simulator to produce a large set of synthetic detections, cloning them as described in Section~\ref{ss:hmag}.
We then randomly draw from this set of synthetic detections until we have a total of 85 synthetic detected 3:2 objects (the number matching our real observational sample). 
The number of non-resonant particles drawn while building up the resonant sample is the number of expected classical detections from $a=38.81-40$ au for OSSOS+.
Figure \ref{fig:cc_sampling_ex} shows one such result of this random sampling procedure.
We repeat this process $10^5$ times to build a distribution of the number of expected detected classical objects for the 1 Gyr and 4.5 Gyr simulation states, and the resulting distribution is shown in Figure~\ref{fig:c_sampling}.
It is clear that the expected number of detected stable classicals near the 3:2 resonance from the 4.5 Gyr simulation snapshot is consistent with the real observed number of objects in the same range.
This serves as an independent verification that the $q=36$ au cut in our simulated phase space is consistent with the observations.

\section{Summary}\label{summary}

We investigate whether the orbital distribution of objects in Neptune's 3:2 mean-motion resonance is consistent with a  history in which orbital phase space was uniformly filled and subsequently ``sculpted" by dynamical stability.  
We find that this simplified model, motivated by dynamical upheaval histories that scattered planetesimal debris outward early in the life of the solar system, is consistent with ensemble data from the Outer Solar System Origins Survey within the uncertainties, with a few notable exceptions. 

Stability sculpting does not substantially alter the inclination distribution of resonant particles, so this distribution must be determined by a different mechanism.  
More subtly, it can be seen in Figure~\ref{fig:cdfs1} that the simulation produces a smaller fraction of objects with mid-high libration amplitudes compared to those observed. 
We suggest that this discrepancy could be due to not accounting for transient populations of objects, which are known to consist of objects that are less deep in the resonance, with higher libration amplitudes \citep[e.g.,][]{Lykawka:2007,yu2018}.
Finally, the fraction of resonant objects in the Kozai sub-resonance is significantly underpredicted in our simulation. 
We find that smooth migration over 1 au at the end of the epoch of planetary upheaval does not alter our model's agreement with the data, but also is not sufficient to push objects into the Kozai portion of the resonance.  Future work is needed to determine whether a longer-distance smooth migration may be accommodated.

We comment that \cite{Pike:2017r} analyze the distribution of test particles throughout the trans-Neptunian region from the \cite{Brasser:2013} simulation of a specific instability model (based on \citealt{Levison2008}) that included Neptune's residual migration from an eccentric orbit at $a=27.5$ au to its current low-eccentricity orbit at 30.1 au. 
\cite{Pike:2017r} find a Kozai fraction in their 3:2 population of 21\%, which is double the Kozai fraction in our simulations.
The libration amplitudes they find for the 3:2 resonant population are also shifted toward slightly higher libration amplitudes compared to our simulations, possibly a result of the high-eccentricity phase of Neptune's orbit, offering an alternative potential origin for the small observed excess of high-libration amplitude objects compared with our model. 

Overall, given the simplicity of our model, we consider the match between the observed population of 3:2 resonant TNOs and our model to be very good, suggesting that stability sculpting likely played a large roll in determining the current distribution of 3:2 resonant objects, particularly in semi-major axis and eccentricity.  We find strong evidence that, if a ``phase-space filling" scattering history provided the initial conditions for this sculpting, the scattering region extended to approximately 36 au.

\section*{Acknowledgements}
RMC, SB, NZ, NH, AHR, JB, JG, and ZS acknowledge support from NSF (grant CAREER AST-1411536/1663706) and NASA (grant NNX15AH59G/NNX17AK64G). KV acknowledges support from NSF (grant AST-1824869) and NASA (grants NNX15AH59G, and 80NSSC19K0785). AHR thanks the LSSTC Data Science Fellowship Program, which is funded by LSSTC, NSF Cybertraining Grant \#1829740, the Brinson Foundation, and the Moore Foundation; her participation in the program has benefited this work. We acknowledge use of the lux supercomputer at UC Santa Cruz, funded by NSF MRI grant AST 1828315. 

\section{Data Availability}\label{data_avail}
The data underlying this article are available in github, at https://github.com/sbalaji718/KBR.

\bibliographystyle{mnras}
\bibliography{references}
\label{lastpage}
\end{document}